# Multi-Objective Service Composition in Ubiquitous Environments with Service Dependencies

Nebil Ben Mabrouk, Nikolaos Georgantas, and Valérie Issarny
Inria Paris-Rocquencourt, Domaine de Voluceau, Le Chesnay, 78150, France.
E-mail: nebil.ben-mabrouk,nikolaos.georgantas,valerie.issarny@inria.fr

**Abstract**

Service composition is a widely used method in ubiquitous computing that enables accomplishing complex tasks required by users based on elementary (hardware and software) services available in ubiquitous environments. To ensure that users experience the best Quality of Service (QoS) with respect to their quality needs, service composition has to be QoS-aware. Establishing QoS-aware service compositions entails efficient service selection taking into account the QoS requirements of users. A challenging issue towards this purpose is to consider service selection under global QoS requirements (i.e., requirements imposed by the user on the whole task), which is of high computational cost. This challenge is even more relevant when we consider the dynamics, limited computational resources and timeliness constraints of ubiquitous environments.

To cope with the above challenge, we present QASSA, an efficient service selection algorithm that provides the appropriate ground for QoS-aware service composition in ubiquitous environments. QASSA formulates service selection under global QoS requirements as a set-based optimisation problem, and solves this problem by combining local and global selection techniques. In particular, it introduces a novel way of using clustering techniques to enable fine-grained management of trade-offs between QoS objectives. QASSA further considers: (i) dependencies between services, (ii) adaptation at run-time, and (iii) both centralised and distributed design fashions. Results of experimental studies performed using real QoS data are presented to illustrate the timeliness and optimality of QASSA.

___________________________________________________________

## 1 INTRODUCTION

Ubiquitous (computing) environments enable integrating and composing, on the fly, services that are offered by (hardware and software) resources available in the environment in order to fulfil complex tasks required by users.

Nevertheless, fulfilling the user's tasks from the functional point of view only is not enough to gain user satisfaction. Users further require a certain Quality of Service (QoS) when exerting their tasks. For this reason, a lot of research efforts in ubiquitous computing have been devoted to the composition of services under the user's QoS requirements, which is known as QoS-aware service composition. QoS-aware service composition is a broad topic. At the core of this topic is the issue of QoS-aware service selection, which allows determining services available in ubiquitous environments and able to meet the user's QoS requirements. The problem arises when dealing with complex user tasks formed of multiple (abstract) activities, and each activity can be achieved using several services that are functionally equivalent, but providing different QoS levels. The question to be asked is then: "*what are the services that should be selected for each activity in the user's task in order to meet the user's QoS requirements and produce the highest QoS?*"

Addressing the above question is even more complicated when considering the challenges entailed by the characteristics of ubiquitous computing. These challenges are mainly about: (i) timeliness (i.e., achieving service composition in a timely manner with respect to on-the-fly interaction with users in ubiquitous environments), (ii) considering dependencies between services, (iii) adaptation support at run-time, and (iv) supporting both centralized and distributed infrastructures in ubiquitous environments. To give a concrete example of ubiquitous environments where such issues may occur, we introduce the following motivating scenario, then we present our solution and the contributions of the paper.

### 1.1 Ubiquitous Shopping scenario

*We introduce a ubiquitous shopping scenario where shopping malls (abstracted as ubiquitous environments) offer smart platforms to assist customers buying their desired items, with respect to a total shopping budget and duration. We specifically consider this scenario in the context of airport shopping malls, where passengers (who want to buy some tax-free items before taking their flights) have strict timeliness constraints.*

*In our scenario, passengers use their mobile devices to submit requests to the shopping platform. The requests specify: (i) the shopping activities (i.e., the set of items to buy and their descriptions), and (ii) QoS requirements such as the total price of the items and the total delivery time.*

*We assume that all shops in the mall propose digital shopping services to advertise the items to sell, their features and their prices, as well as the QoS capabilities of*

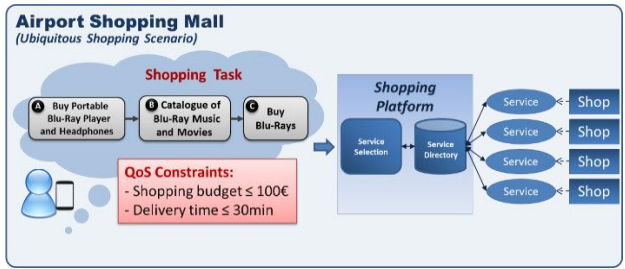

*Figure 1. Ubiquitous shopping scenario*

shops, e.g., delivery time. All services are supposed to be published in a directory within the shopping platform.

The platform provides to passengers several compositions of shopping services that offer the required items and meet their QoS requirements. The proposed compositions are ranked according to their overall QoS (i.e., total price and delivery time).

To make our scenario even more challenging, we consider the shopping task in an ad hoc context, such as open-air markets, where there are no centralised shopping platforms to assist customers. Such environments are rather fully distributed and formed of mobile and resource-constrained devices. Vendors and customers user their mobile devices to advertise their services and fulfil their shopping tasks, respectively.

To address the challenges illustrated in the above scenario, we introduce an efficient service selection algorithm called QASSA (Qos-Aware Service Selection Algorithm). Next, we present the contributions of QASSA and give the outline of the paper.

## 1.2 Contributions of the paper

QASSA presents the following contributions:
1) It models service selection under global QoS requirements as a set-based multi-objective optimisation, benefiting from recent mathematical proposals [1];
2) QASSA resolves QoS-aware service selection efficiently using clustering techniques, and distinguishes several classes of services that represent different trade-offs between QoS properties, hence enabling fine-grained management of these trade-offs with respect to user preferences;
3) QASSA is dependency-aware, i.e., it considers dependencies between services when performing the selection;
4) QASSA selects several alternative service compositions (instead of only one), thus enabling service substitution (and accordingly adaptation support) at run-time;
5) QASSA is devised in both centralised and distributed fashions, which makes it suitable for both resource-enabled and resource-limited ubiquitous environments;
6) QASSA executes in a timely manner (on top of both resource-enabled and resource-limited devices) while achieving a near-optimal QoS (further details are given in Section 4);

The paper is structured as follows. In Section 2, we present the design rationale of QASSA, its underlying problem definition, as well as its both phases, viz., local and global selection phases. After that, we introduce some enhancements to QASSA that address respectively service dependencies, adaptation at run-time and distributed design (Section 3). Finally, we give the results of an extensive experimental study illustrating the efficiency of QASSA in terms of timeliness and optimality (Section 4), and we conclude in Section 6.

## 2 THE QASSA ALGORITHM

### 2.1 Notations and Basic Assumptions

QASSA is initialised by taking as input a user request R, which is defined as a quadruple $R = (T, U, P, W)$, where $T$ refers to the required task and $U$ refers to global QoS constraints $U = \langle u_1, \dots, u_n \rangle$ imposed by the user on a set of QoS properties $P = \langle p_1, \dots, p_n \rangle$. For each constraint, the user specifies the relative importance of its associated QoS property by giving a set of weights $W = \langle w_1, \dots, w_n \rangle$ where $w_i$ is the weight of QoS property $p_i$. It is worth noting that the sum of all the weights must be equal to 1, i.e., $\sum_1^n w_i = 1$.

Several approaches in the literature (e.g., [32]) address the way users define their QoS preferences.

For a user task T, its structure is specified as a set of activities $T = \langle A_1, \dots, A_z \rangle$ coordinated by execution patterns (e.g., sequence, parallel, and loop), further details about the used composition patterns are given in [2]. To each activity $A_i$ in $T$ is associated a set of concrete service candidates $S = \{s_{i,1}, s_{i,2}, \dots, s_{i,m_i}\}$ that are able to realise $A_i$. We consider stateless services that can be bound to one or more abstract activities in the composition. Additionally, service compositions are specified such that the control flow and data flow are intertwined, i.e. messages that represent a service request (control flow) also hold the input data –if any- for the service they trigger (data flow). Each service $s_{i,k}$ ($1 \leq k \leq m_i$) (associated with an abstract activity) is represented by its QoS vector $QoS_{s_{i,k}} = \langle q_1, \dots, q_n \rangle$, where $q_j$ is the advertised value of the QoS property $p_j$ ($1 \leq k \leq n$). QoS values are supposed to be specified by service providers based on previous executions of the services (they are increasingly included in service level agreements).

The overall QoS of service compositions is function of the execution patterns structuring the composition, and it determined with respect to three QoS aggregation approaches: (1) best-case approach (i.e., considering the best QoS value), (2) worst-case approach (i.e., considering the worst QoS value), and (3) mean-value approach (i.e., considering the average of services' QoS values). Further details about QoS properties, preferences and QoS aggregation formulae are given in [2].

## 2.2 Design Rationale

QoS-aware service selection algorithms fall under two broad classes with respect to their selection techniques. On the one hand, *local selection* (i.e., greedy selection) proceeds by selecting the best service in terms of QoS for each activity in the user task separately. This technique has a low computational cost but it cannot guarantee meeting global QoS requirements. On the other hand, *global selection* covers the scope of the whole composition and ensures meeting global QoS requirements. However, it is of high computational complexity.

The combination of local and global optimisation is a general and powerful technique to extract optimal compositions in diverse scenarios [3]. Accordingly, QASSA combines local and global selection techniques with respect to the bi-level optimisation model [4]. This model is defined as a hierarchy of two optimisation problems (upper-level or leader, and lower-level or follower problems). Each problem is optimised separately without considering the objective of the other one. However the decision made at the upper-level affects the objective space of the lower-level as well as the decision space.

In accordance with the bi-level optimisation model, QASSA proceeds through two main steps: (1) *local selection* (representing the upper-level optimisation problem), which aims at selecting services with the highest QoS for each activity in the user task, and (2) *global selection* (representing the lower-level optimisation problem), which aims at selecting near-optimal compositions of services resulting from the local selection. QASSA further selects several alternative near-optimal compositions. Indeed, selecting only one service composition brings about several shortcomings such as the lack of choices for the user, the overload of hot services (i.e., services with high QoS) [5], and the lack of adaptation support [6] (i.e., deferred final selection and dynamic binding at run-time).

## 2.3 Local Selection Phase

Most of existing QoS-aware service composition approaches use utility functions to evaluate the overall QoS of the composition [7]. Such functions generally: (i) scale QoS values of single properties, (ii) set weights to the corresponding properties based on user preferences, then (iii) sum up the properties multiplied by the weights. Thus, they reduce multi-objective QoS-aware service composition to a single objective optimisation, which brings about the issue of balancing low values of one or more QoS properties by good values of other properties.

In QASSA, we focus on providing a mechanism that enables fine-grained management of the trade-offs between QoS objectives. We express the local selection as a multi-objective optimisation problem, and we aim at solving this problem while producing several solutions (and not a single solution), thus enabling the global selection phase and adaptation support at run-time.

### 2.3.1 PROBLEM DEFINITION

A recent proposal by Zitzler et al. [1] defines multi-objective optimisation as a set problem having as goal to identify the Pareto optimal set (i.e., the best solution set) among several solution sets, each set reflecting a specific trade-off between the optimisation objectives [8]. Determining the Pareto optimal set is typically exponential in the size of the problem instance [9]. Therefore, resolving set-based multi-objective optimisation is often reduced to identifying a good Pareto set approximation.

In accordance with Zitzler's proposal, we define the local selection problem as a set-based multi-objective optimisation. Consider the optimisation of the QoS vector $P = \langle p_1, \dots, p_n \rangle : S \to \mathbb{R}^n$ where all QoS properties $p_i$ are, without loss of generality, to be maximised. Here, $S$ denotes the feasible set of solutions, i.e., the set of service candidates of a given activity in the user task. A single service $s \in S$ will be denoted as a decision vector or solution $(s = \langle q_1, \dots, q_n \rangle)$ where $q_i = p_i(s)$.

We define the search space $\mathbb{S}$ by the set of feasible sets of services (and not single services). In the context of QoS optimisation, an element (i.e., a service-set) in $\mathbb{S}$ is called *QoS Class*, and denoted $QC \in \mathbb{S}$. A QoS class represents a set of services having *roughly* the same QoS and reflecting the same trade-off between QoS properties.

To enable the comparison of QoS classes, a set preference relation must be defined over $\mathbb{S}$. A set preference relation provides the information on the basis of which the selection is carried out; it says whether a QoS class is better (in terms of QoS) than another one, or not. The set preference relation can be defined in terms of a *quality indicator* (such as the hypervolume indicator [9]). The quality indicator is a function that assigns, to each solution-set, a scalar value reflecting its quality according to a particular goal, i.e., a fitness function defined over sets. In this paper, we need to define a quality indicator $I_q$ that is specific to our QoS optimisation problem (see Equation 2). Our objective is then to find a solution-set that maximises the value of $I_q$ as defined below (the operator *argmax* returns the QoS class for which $I_q$ attains its maximum value):

$$argmax\left(I_q(QC)\right) \quad \text{where } QC \in \mathbb{S} \quad (1)$$

Solving this problem consists in determining $\mathbb{S}$ and its underlying QoS Classes $QC \in \mathbb{S}$, as well as defining the quality indicator $I_q$. Towards this purpose, we propose investigating clustering techniques, notably the K-means algorithm. Clustering techniques allow for grouping a set of data into several clusters with respect to given criteria. If we apply the same principle to our purpose (i.e., QoS-aware service selection), we can group service candidates associated with an activity into several clusters according to their QoS values. Each cluster includes services having roughly the same QoS. We can further define a quality indicator on these clusters of services based on their respective QoS properties. Next, we show how to solve the local selection problem (defined as a set-based multi-objective optimisation) using K-means, further introducing the formal definition of QoS Class and quality indicator.

### 2.3.2 LOCAL SELECTION IN QASSA

To select a set of services providing high values for all QoS properties, we perform one-dimensional clustering applied *n* times (once per QoS property). That is, we cluster service candidates (i.e., associated with each activity in the composition) for each QoS property separately. Thus, for each QoS property $p_j$ we obtain several clusters $C = \langle c_{1,j}, \ldots, c_{g,j} \rangle$, going from the cluster $c_{1,j}$ of services having the lowest values of $p_j$ to the cluster $c_{g,j}$ of services with the highest values for the same property (respecting the definitions of positive and negative QoS properties). After that, by considering the intersection of the clusters with the highest values associated with each QoS property, we obtain the service-set providing high values for all QoS properties jointly. To formally explain the local selection in QASSA, we introduce the concept of *QoS Level* and *QoS Class*.

The concept of QoS Level is used to group together service clusters having roughly the same quality level for all QoS properties. The number of QoS levels corresponds to the number of clusters for each QoS property denoted g. For instance, in Figure 2 we cluster candidate services into 3 clusters, thus obtaining 3 QoS levels (1, 2 and 3) corresponding to the clusters of services having respectively 'low', 'medium' and 'high' QoS values for all QoS properties. As we have 4 QoS properties, each QoS level comprehends 4 clusters. Below, we formally define the QoS level concept.

*Definition 1:* Given a set of QoS properties $P = \langle p_1, \ldots, p_n \rangle$ and a set of services $S = \langle s_{i,1}, s_{i,2}, \ldots, s_{i,m_i} \rangle$ associated to activity $A_i$ and grouped into $g$ clusters $\langle c_{1,j}, \ldots, c_{g,j} \rangle$ for each QoS property $p_j$ (where $c_{1,j}$ is the cluster of services with the lowest values of $p_j$ and $c_{g,j}$ is the cluster of services with the highest values of $p_j$), we define a QoS level $QL_l = \{c_{l,1}, \ldots, c_{l,n}\}$ as the set of clusters

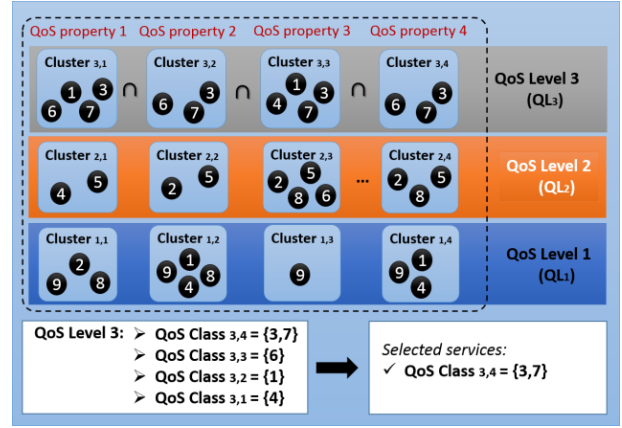

*Figure 2. Local selection in QASSA*

associated with each QoS property $p_j$ and having the same level $l$ $(1 \leq l \leq g)$.

As stated above, a QoS level $QL_l$ is used to group clusters with the same level $l$ together, thus we can perform their intersection and determine services with QoS values in this level. In particular, we are interested in the best QoS level $QL_g = \{c_{g,1}, \ldots, c_{g,n}\}$ which groups clusters with the highest QoS values. The intersection of these clusters yields services with the highest QoS values for all QoS properties. However, if the intersection produces an empty set, we investigate other combinations (i.e., intersections) of clusters within $QL_g$. To do so, we introduce the concept of QoS Class, which represents different intersections of clusters within a given QoS level. The QoS class concept is formally defined as follows.

*Definition 2:* Given a QoS level $QL_l = \{c_{l,1}, \ldots, c_{l,n}\}$, we define a QoS class $QC_{l,e}$ $(1 \leq e \leq n)$ as the intersection of $e$ clusters among $QL_l$. Consequently, a QoS level $QL_l$ comprehends several QoS classes (e.g., $\{QC_{l,1}, \ldots, QC_{l,n}\}$).

Literally, a QoS class $QC_{l,e}$ represents the set of services having exactly $e$ QoS properties out of $n$ at the QoS level $l$. According to this, the QoS class $QC_{g,n}$ groups the best set of services in terms of QoS, since they have all their $n$ QoS properties in the highest QoS level $QL_g$. If $QC_{g,n}$ is an empty set (i.e., there are no services with high values for all QoS properties), we try to find the next best QoS class in terms of QoS (e.g., $QC_{g,n-1}$).

Nevertheless, we may obtain several QoS classes having the same level and the same number of QoS properties (e.g., selecting $n-1$ QoS properties out of $n$). To determine the best QoS class, we use the quality indicator $I_q$ (already introduced in Section 2.3.1). $I_q$ is defined based on the level $l$ and the number of QoS properties $e$ of the considered QoS class, as well as the weights $w_j$ associated with these QoS properties. That is, the quality indicator $I_q$ of the QoS class $QC_{l,e}$ is higher (i.e., it includes services with better QoS) when: (i) it is associated with a QoS level $QL_l$ of a higher level $l$, (ii) it comprehends a higher number of QoS properties $e$ in that level, and (iii) the weights $w_j$ associated with these QoS properties are more important for the user. $I_q$ is formally defined as follows:

$$I_q(QC_{l,e}) = l \times e \times \sum_{j=1}^{e} w_j, \quad w_j \in \{W : c_{l,j} \in QC_{l,e}\} \quad (2)$$

Finally, it is worth noting that QASSA implements the local selection using a variant of K-means called K-means++ [10], which takes as input only the number of clusters (in opposition to K-means which takes as input the number of centroids and their initial coordinates).

The number of clusters is determined beforehand by learning from the previous executions of the considered activity using existing techniques in the literature (viz., the Davies-Bouldin index [11]) that determine whether a given integer $g$ is the most appropriate number of clusters for classifying a fixed data set. The Davies-Bouldin index mainly uses the distance separating the clusters as a metric for evaluating $g$. The overall execution of the local selection phase of QASSA is described in Algorithm 1.

## 2.4 GLOBAL SELECTION PHASE

The global selection phase aims at composing locally selected services and determining near-optimal service compositions, i.e., service compositions that: (i) satisfy the global QoS requirements, and (ii) maximise the QoS offered to the user. In our approach, we focus on selecting several alternative service compositions (and not a single composition).

When dealing with global optimisation problems with multiple objectives and a large number of potential solutions, heuristic algorithms are the only possible choice [12]. We focus on using population-based heuristics (e.g., genetic algorithms (GA), differential evolution (DE) algorithms [13]) to solve QoS-aware service selection under global QoS requirements. This class of algorithms consists in recursively optimising an initial solution using operators such as crossover/mutation. Starting from the fact that our local selection approach is highly selective in the sense that it selects few services having a high QoS level (which reduces considerably the number of services to be investigated), we argue that a population-based heuristic can quickly produce near-optimal service compositions.

In accordance with the above, we adopt a population based approach to solve the global selection phase of QASSA, viz., the Controlled Random Search (CRS) algorithm, which is a population-based global optimisation heuristic like GA and DE. The choice of CRS is motivated

```
input  : A set of activities T = {A_1, ..., A_z} of size z;
         A set of services S_i = {s_{i,1}, ..., s_{i,m_i}} of size m_i for each activity A_i
         (i ∈ {1, ..., z});
         A set of QoS properties P = {p_1, ..., p_n} of size n;
         A set of weights on QoS properties W = {w_1, ..., w_n} of size n;
         A QoS vector QoS_{s_{i,k}} = ⟨q_1, ..., q_n⟩ for each service s_{i,k} (k ∈ {1, ..., m_i}).
output: A QoS class for each activity A_i
begin
    foreach A_i ∈ T do
        // Services' clustering
        foreach p_j ∈ P do
            (Step 1) Apply K-means++ to group the services S_i into g
                clusters w.r.t. their values for p_j
            K-means++(S_i, g) ⇒ {c_{1,j}, ..., c_{g,j}};
        end
        // Services' selection
        for l = g downto 1 do
            for e = n downto 1 do
                (Step 2) Build QL_l as the set of n clusters of level l for
                    all the QoS properties
                QL_l = {c_{l,1}, c_{l,2}, ..., c_{l,n}};
                (Step 3) For each combination C^e_{|QL_l|} of e clusters
                    among those of QL_l
                foreach C^e_{|QL_l|} do
                    (Step 4) Build QoS class e of the level l as the
                        intersection of the clusters belonging to C^e_{|QL_l|}
                    QC_{l,e} = {s_{i,k} : s_{i,k} ∈ ∩ C^e_{|QL_l|}};
                    (Step 5) Initialise the QoS class to be selected and
                        its quality indicator
                    Result = ∅;
                    Result_I_q = 0;
                    (Step 6) Select the QoS class with the highest
                        quality indicator
                    if QC_{l,e} ≠ ∅ then
                        I_q = r × e × ∑_{j=1}^{y} w_j;
                        if I_q > Result_I_q then
                            Result_I_q = I_q;
                            Result = QC_{l,e};
                        end
                    end
                end
            end
        end
        return Result;
    end
end
```

*Algorithm 1. The local selection algorithm*

by the fact that it is a simple algorithm capable of global optimisation, subject to inequality constraints [14]. CRS initially builds a preliminary service composition by selecting a service (among those resulting from the local selection phase) for each abstract activity in the user task. The global QoS of the composition is then computed with respect to the structure of the composition and QoS aggregation formulae (as detailed in our previous work [15]. If the global QoS meets the user requirements, the composition is then considered as a solution. The global QoS of the composition is then gradually enhanced by randomly replacing a single service with an alternative one. Accordingly, QASSA yields as a result several alternative service compositions ranked with respect to their overall QoS.

## 2.5 COMPUTATIONAL COMPLEXITY ANALYSIS

We analyse the computational complexity of the local and global selection phases of QASSA. Concerning the local selection phase, we do not consider the complexity of computing the Davies-Bouldin index (which is used to decide about the number of clusters, see Section 2.3.2), since the computation is performed off-line, i.e., before executing QASSA.

As already explained, the local selection phase is performed using the K-means++ algorithm, the complexity of which is of $O(\log(K))$ [10] where $K$ denotes the number of clusters. As we cluster services (i.e., execute K-means++) for each QoS property and for all the activities in the user task, the overall complexity of local selection is then of $O(Z.N.\log(K))$ where $Z$ and $N$ denote the number of activities in the user task and the number of QoS properties, respectively. Therefore, the local selection phase runs in a linearithmic time.

Concerning the global selection phase, its computational complexity can be determined based on the fact that we proceed similarly to the CRS algorithm [14]. That is, when iteratively checking service compositions, we replace a single service per composition in each iteration. In accordance with this, we first compose $Z$ services (one service per activity) to build the initial service composition, then we iterate on checking the remaining $T-Z$ services (one service per iteration), where $T$ denotes the total number of services associated with all the activities in the user task. Therefore, the total number of compositions to check (i.e., more specifically the number of iterations) is $T-Z+1$. Additionally, for each composition, we execute $Z.N$ arithmetic instructions to aggregate the $N$ QoS values of the $Z$ services forming the composition. Then, we execute $N$ comparison instructions to determine whether the $N$ QoS values of the composition satisfy the global QoS requirements of the user. The global selection phase runs then in quadratic time of $O((Z.N+N)(T-Z+1))$.

Based on the above results, we state that QASSA executes in quadratic time, thus it reduces considerably the computational complexity of service selection under global QoS requirements, known to be NP-hard [16].

## 3 ENHANCING QASSA

We endow QASSA with three important capabilities that respectively address: (i) service dependencies, (ii) adaptation at run-time and (iii) distributed design. Below, we briefly explain the importance of each capability and how we achieve it.

## 3.1 MANAGING SERVICE DEPENDENCIES

Service composition implies coordination between the underlying services, which may entail dependencies between these services. Service dependencies can considerably impact the selection and composition process, as well as the overall obtained QoS. Accordingly, we aim at making QASSA dependency-aware, i.e., it takes into account service dependencies during the selection process, thus ensuring that every selected service is compatible with other ones. Service dependencies can be either defined by users when expressing their desired tasks or automatically generated using existing approaches in the literature (e.g., [31]).

### 3.1.1 SERVICE DEPENDENCY CLASSES

QASSA considers two broad classes of service dependencies called intra-dependencies and inter-dependencies [17]:

- *Intra-dependencies* occur when two or more abstract activities in the user task must be accomplished by the same service. This is particularly true for coarse-grained services that can support multiple activities in the user task. Intra-dependencies are generally due to interoperability, QoS and business reasons. For instance, binding the same service to multiple activities in the user task makes service coordination easier or may enhance the overall QoS of the composition. To give a concrete example of intra-dependencies, we recall our motivating scenario where the shopping task required by the customer is composed, e.g., of three activities: (A) buying a portable Blu-ray player and headphones, (B) getting the catalogue of Blu-ray music and movies, and (C) buying Blu-rays (see Figure 3.1). The activities B and C should be correlated with an intra-dependency, since the Blu-rays must be bought from the same shop providing the catalogue, otherwise interoperability issues may arise.

- *Inter-dependencies* concern correlations between separate services associated with two or more activities in the user task. They include various types of correlations, notably [18]: (i) Input/Output dependencies (i.e., a service requires/or provides data from/to another service), (ii) Cause/Effect dependencies (i.e., a service has preconditions to be satisfied based on the effect of other services), and (iii) User-constraint dependencies (i.e., dependencies imposed by users on specific services). The mentioned types of inter-dependencies concern jointly the data and control flows. As already introduced, in QASSA the control flow and data flow are intertwined. To give an example of inter-dependencies, we continue with our example of the shopping task. In this example, service candidates of the activity A should have Cause/Effect inter-dependencies with those of activity B, since the latter activity cannot be accomplished unless the former one is achieved. That is, getting the catalogue and buying Blu-rays is useless if the customer does not buy a Blu-ray player and headphones.

At a global view, intra-dependencies can be seen as correlations between abstract activities, whereas interdependencies concern correlations between concrete services. This classification impacts the way service dependencies are handled in QASSA, which will be explained below.

### 3.1.2 PRE-PROCESSING SERVICE DEPENDENCIES

QASSA deals with the above service dependencies at a preliminary step that takes place before proceeding to services' selection. The objective of QASSA is to combine the abstract activities concerned by the dependencies into a single coarse-grained activity, then process it through the local and global selection phases as already explained. QASSA manages each kind of the above service dependencies in a different manner, as depicted in Figure 3.

Abstract activities correlated with intra-dependencies are merged into a single coarse-grained activity having as candidate services the intersection of those services associated with the considered activities (i.e., services that are common to these activities, thus they are able to fulfil them jointly). Figure 3.2 illustrates the operation of merging the activities B and C (correlated with an intra-dependency), which have two common services $s_1$ and $s_3$. The result of this operation is a coarse-grained activity called BC with $s_1$ and $s_3$ as candidate services.

Abstract activities concerned with inter-dependencies are also combined into a single coarse-grained activity, however QASSA introduces new fictive services as candidates to fulfil this activity. The newly introduced services are coarse-grained as they comprehend two or more services connected via inter-dependencies. Indeed, for each inter-dependency linking two services $s_x$ and $s_y$, QASSA creates a fictive service called $S_{xy}$, having as QoS the aggregation of QoS values of $s_x$ and $s_y$. The details of QoS aggregation are given in our previous work [15]. Figure 3.3 depicts an example of managing inter-dependencies between two abstract activities A and B.

When dealing with complex service correlations formed of both kinds of dependencies (i.e., intra-dependencies and inter-dependencies). QASSA proceeds by managing at first intra-dependencies (i.e., keeping only services that are common to the activities linked via intra-dependencies), then it handles inter-dependencies with respect to the procedure explained above. Figure 3.4 shows an example of managing three activities A, B and C, with inter-dependencies between A and B, and an intra-dependency between services of B and C.

Once service dependencies are pre-processed, QASSA performs the local and global selection phases as already introduced. If some fictive services are selected (i.e., they make part of the resulting compositions), they are simply replaced by their initial values (their associated concrete services) before executing the user task.

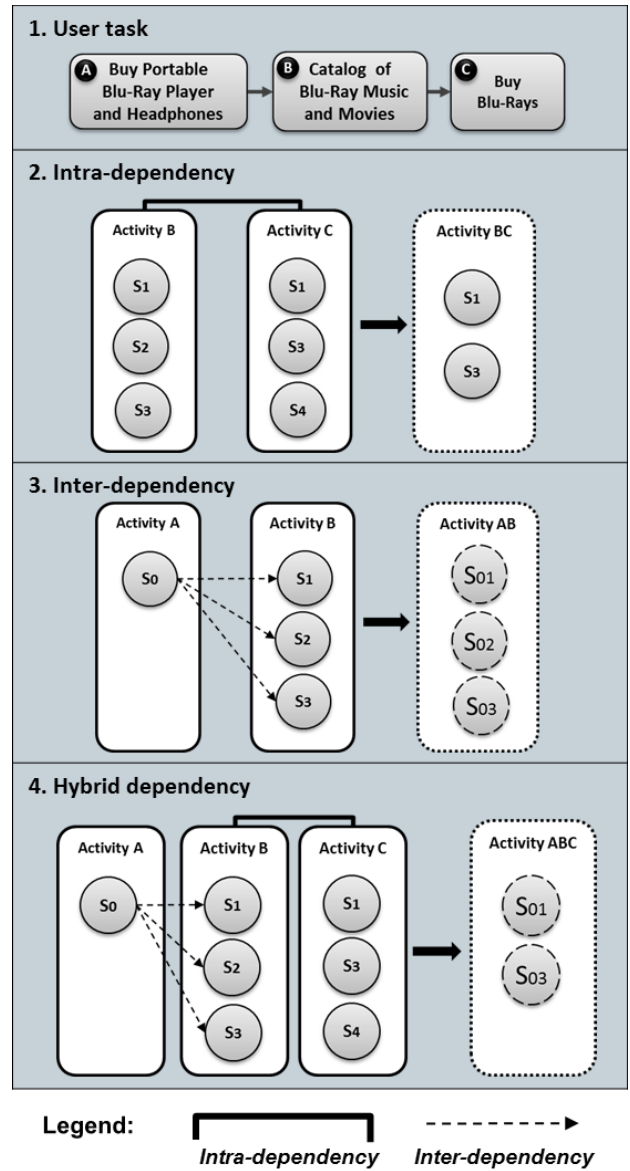

*Figure 3. Illustrating service dependencies*

## 3.2 COMPOSITION ADAPTATION

Service compositions have to be adapted at run-time with respect to the dynamics of ubiquitous environments, notably QoS fluctuations. By adaptation we refer to the ability to alter service compositions in response to changes impacting their executions [2]. The adaptation process is triggered when one or more running services show a faulty behaviour, i.e., they are no longer available, fail or provide unsatisfactory QoS. For instance, if we recall our motivating scenario, some requested items may be no longer available at run-time (they are sold), or the delivery time of the shopping services may increase depending on the number of customers and the delivery ability of shops. Hence, the shopping platform must adapt service compositions dynamically at run-time, while always meeting QoS requirements of the customer. To adapt a running service composition, QASSA proceeds by substituting one or more services based on those initially selected during the local and global selection phases. The goal of service substitution is to maintain the level of QoS uppermost with respect to the threshold defined by user constraints, while taking into account service dependencies.

Two strategies of service substitution are possible to this regard:

- *Substituting single services*: It consists in replacing each faulty service in the composition with an alternative one among those previously selected. More specifically, QASSA investigates services associated with the same abstract activity as the faulty service, and chooses the service yielding the best overall QoS (for the whole composition). If the faulty service has some dependencies with other ones, QASSA performs the substitution based on the coarse-grained services established during the pre-processing phase of service dependencies.

- *Substituting a sub-composition*: Whenever the strategy of substituting single services fails (due to QoS violation or absence of services), QASSA attempts to substitute the remaining sub-composition, i.e., the sub-composition not executed yet. QASSA investigates service compositions previously selected and attempts to find an alternative sub-composition that allows to achieve the user task.

When both substitution strategies fail, QASSA asks the user to relax one or more QoS requirements, and proceeds through the local and global phases to find alternative service compositions with respect to the new requirements.

## 3.3 DISTRIBUTING QASSA

The version of QASSA presented in Section 2 assumes the presence of a centralised resource-enabled infrastructure supporting QoS-aware service composition. Nevertheless, in ubiquitous environments, it is not always possible to assume the support of such an infrastructure. QoS-aware service composition in ubiquitous environments can be rather underpinned by *ad hoc* infrastructures formed of mobile and resource-constrained devices. For instance, as already mentioned in our motivating scenario, the shopping task of customers can take place in open-air markets with no centralised infrastructures. For this reason, we present a distributed version of QASSA, which is capable of operating on top of *ad hoc* infrastructures.

Distributed QASSA enables accomplishing service selection as a synergistic interaction between the user device (referred to as requester) and other devices available in the

```
input  : A set of activities T = {A₁,..,A_z} of size z;
         A set of services S_i = {s_{i,1},..,s_{i,m_i}} of size m_i for each activity A_i
         (i ∈ {1,..,z});
         A set of QoS requirements U = {u_1,..,u_n} of size n;
         A set of weights on QoS requirements W = {w_1,..,w_n} of size n;
         A QoS vector QoS_{s_{i,k}} = ⟨q_1,..,q_n⟩ for each service s_{i,k} (k ∈ {1,..,m_i}).
output: A set of service compositions.
begin
    (Step 1) Broadcasting help message and getting helpers
    broadcast (help message);
    foreach device d favourably replying to the help message do
        helpers = helpers ∪ d;
    end
    (Step 2) Splitting the user request into a set E of elementary requests
    foreach d ∈ helpers do
        e_i = ⟨{(A_x, S_x),..,(A_y, S_y)}, U, W⟩;
        E = E ∪ e_i;
    end
    while E ≠ ∅ do
        (Step 3) Scheduling elementary requests to helpers
        foreach e_i ∈ E do
            e_i → d    (d ∈ helpers);
        end
        (Step 4) Fulfilling the local selection phase
        while Timeout Session do
            Helper Side : execute the local selection (Algorithm 1) given
            e_i as input;
        end
        (Step 5) Getting the results of elementary requests
        foreach e_i ∈ E do
            if result (e_i) ≠ null then
                Selected services for A_i = result (e_i);
            end
            else Break;
            E = E \{e_i};
        end
    end
    (Step 6) Fulfilling the global selection phase
    execute the CRS algorithm given the selected services for T;
end
```

*Algorithm 2. Overview of distributed QASSA from the requester point of view. The coloured box concerns the execution at the helper side.*

environment (referred to as helpers). We assume that a lightweight middleware implementing distributed QASSA is already installed on the requester and helpers' devices. As described in Algorithm 2, the main idea of distributed QASSA is to perform the local selection phase using several helpers simultaneously. That is, we propose to divide the local selection (for the whole user task) into several elementary requests, each dealing with a single abstract activity. Then, elementary requests are flexibly assigned to helpers (with respect to the number of helpers and their computational capabilities). Ideally, for each abstract activity in the user task, the local selection is executed using a separate helper. After that, the requester collects the local selection results from helpers and performs the global selection phase on the user device.

The global service selection is difficult to carry out in a distributed way because it requires a global vision of QoS information and the structure of the composition [19]. Additionally, it typically requires a resource-rich device, given the computational complexity of the problem. As detailed in Section 2.5, our global service selection approach has a low computational complexity; thus it can be carried out using only the resource-constrained device of the requester. The timeliness of our distributed algorithm is further validated by experimental results detailed in the next section.

Table I. Experimental set up

| Centralized QASSA | Distributed QASSA |
|---|---|
| - Machine: Dell<br>- Processor: AMD Athlon 1.80GHz<br>- RAM: 1.8 GB<br>- OS: Windows XP<br>- Programming language: J2SE 1.6 | - Machine: HTC Desire<br>- Processor: Qualcomm QSBD8250 1GHz<br>- RAM: 576 Mo<br>- OS: Android 2.2 (Froyo)<br>- Programming language: Android SDK 2.2 (based on J2SE 1.5) |

## 4 EXPERIMENTAL EVALUATION

We conducted a set of experiments to assess the centralised and distributed versions of QASSA. Table 1 describes the experimental set up used in our experiments. We use a basic setup (with limited computational and memory resources) that can be readily assumed in the context of ad hoc ubiquitous environments. For the evaluation of QASSA, we are interested in two metrics:

- *Execution time:* It measures the timeliness of QASSA with respect to the size of the selection problem in terms of the number of activities and the number of candidate services per activity.
- *Optimality:* It measures how optimal is the QoS utility provided by QASSA. This is determined by the ratio of the QoS utility resulting from QASSA over the optimal QoS utility given by a brute-force algorithm. The optimality metric is then given by the following formula:

$$Optimality = F/F_{opt} \qquad (3)$$

where $F$ is the QoS utility given by our heuristic algorithm, and $F_{opt}$ is the optimal QoS utility given by the IBM ILOG CPLEX Optimiser[1].

For the purpose of our experiments, we focus on the size of service compositions (i.e., the number of activities and the number of services per activity) as well as the used QoS data. We do not consider the functional aspect of service compositions as it does not impact the performance of QASSA. As we do not have real service compositions corresponding to our purpose, we developed a *Composition Generator*, which randomly generates service compositions used for experimenting QASSA. *Composition Generator* takes as parameters the number of activities (denoted $a$) and the number of candidate services per activity (denoted $k$), and it proceeds through two steps: (i) constructing an abstract service composition which comprehends $a$ activities structured with respect to randomly chosen composition patterns, (ii) binding $k$ concrete services to each activity in the composition. QoS values associated with these services are acquired from the QWS dataset available online[2]. This dataset consists of 5000 real Web services, each with a set of 9 QoS properties measured using commercial benchmark tools [20]. Further details about the implementation of QASSA are given in [2].

Once service compositions are generated, we further need to configure the execution of QASSA with respect to the following parameters:

1) *Aggregation approach:* As already introduced in Section 2.1, our algorithm supports three QoS aggregation approaches: worst-case, best-case and mean-value. We opt for the worst-case approach as the default method for aggregating QoS values. We further perform experimentation with respect to the three aggregation approaches in order to study their impact on the timeliness and optimality of QASSA.

2) *Global QoS constraints:* QASSA requires as input global QoS constraints imposed by the user on the whole composition. As we do not have real user requirements, we opt for a statistical method to determine global QoS constraints. For each QoS property (e.g., response time) we calculate the mean value $m_i$ of service candidates associated with each activity $A_i$, then we aggregate all mean values (i.e., $m_1, m_2, ..., m_n$) with respect to the structure of the composition. That is, we set the global QoS constraint of each QoS property to the aggregated mean values (of service candidates) associated with each activity. We further vary the global QoS constraints with respect to different statistical values in order to analyse their impact on the timeliness and optimality of QASSA. Further details are given in Section 4.1.2.

### 4.1 PERFORMANCE OF QASSA (THE CENTRALISED VERSION)

In this section, we present the experimental evaluation of the centralised version of QASSA (using the experimental setup detailed in the left column of Table 1). For the sake of precision, we execute each experiment 20 times and we calculate the mean value of the obtained results.

Figure 4 (a) depicts the execution time of QASSA with respect to the number of services per activity. We fix the number of QoS constraints to 5, vary the number of activities between 10 and 50, and vary the number of services per activity between 50 and 200. The obtained results show that the execution time of our algorithm

---

[1] http://www-01.ibm.com/software/integration/optimization/cplex-optimizer/about/?S_CMP=rnav

[2] http://www.uoguelph.ca/~qmahmoud/qws/index.html

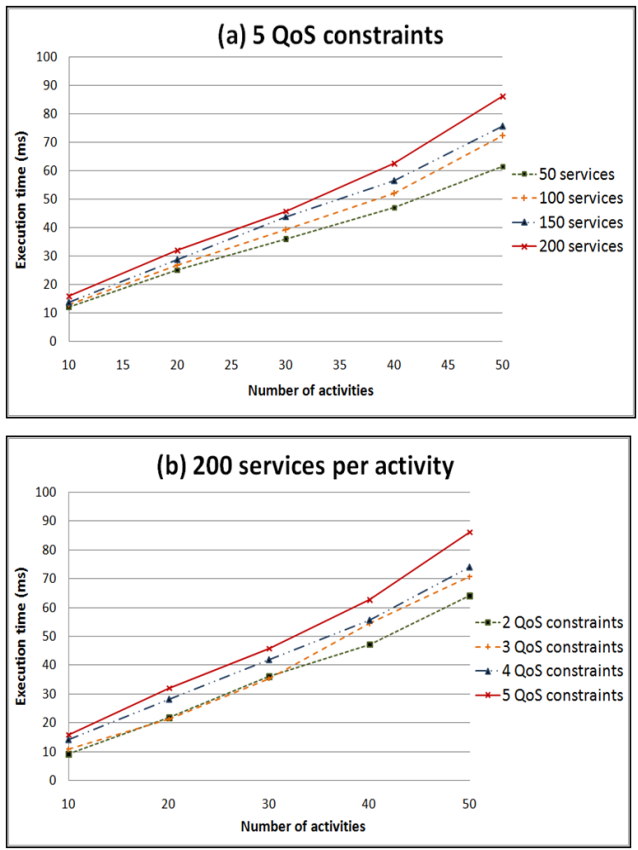

*Figure 4. Execution time while varying (a) the number of services per activity, and (b) the number of QoS constraints*

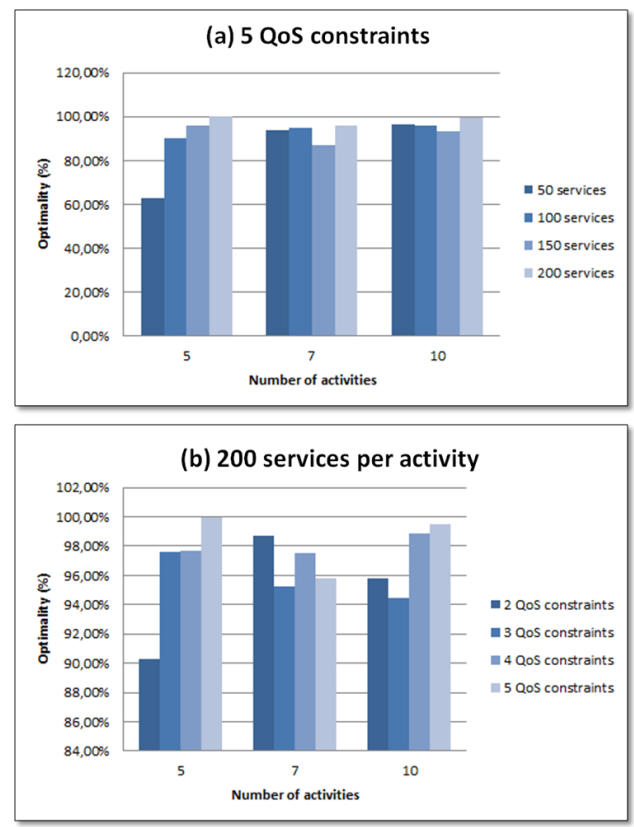

*Figure 5. Optimality measurements while (a) varying the number of services, and (b) the number of QoS constraints*

increases (up to 89ms) along with the number of services, which is an expected result.

Figure 4 (b) depicts the execution time of QASSA with respect to the number of QoS constraints. We fix the number of services per activity to 200 and vary QoS constraints between 2 and 5. The obtained results show that the execution time of our algorithm increases (up to 89ms) along with the number of QoS constraints, which is also an expected result, i.e., a higher number of QoS constraints requires more computational effort, hence a longer execution time.

Both figures show that the execution time of our algorithm increases almost linearly along with the number of activities in the composition. In general, our algorithm executes in a timely manner (i.e., less than 0.09s) with respect to spontaneous interaction with users aimed at by ubiquitous computing. Indeed, guidelines for response time in interactive applications specify that 1s is the limit to keep the user's flow of thought seamless [21].

To have a more accurate idea about the efficiency of QASSA in terms of timeliness, we compare the above obtained results to those published in [16], which presents an efficient approach that combines local and global selection techniques for service selection under global QoS constraints. The authors consider the same dataset (i.e., QWS dataset) and configuration as in our experiments; however they use an experimental setup (a HP ProLiant DL380 G3 machine with 2 Intel Xeon 2.80GHz processors and 6 GB RAM) that is more powerful than the one used to evaluate QASSA. In spite of that, our algorithm achieves the same execution time as [16] (between $\approx 10$ ms and $\approx 90$ ms).

Concerning the optimality of QASSA, we measure it while varying the number of activities between 5 and 10. Figure 5 (a) depicts the optimality of QASSA while fixing the number of QoS constraints to 5, varying the number of activities between 5 and 10 and varying the number of services per activity between 50 and 200. It shows that the optimality of QASSA is generally more than 90%, and it can reach 100%. However, for the specific case of 5 activities and 50 services per activity, the optimality decreases to 60%, which can be explained by the fact that when the number of services decreases, the probability to find services with a satisfactory value for all QoS properties decreases also, hence yielding a low optimality.

Additionally, we measure the optimality of QASSA while fixing the number of services to 200, varying the number of activities between 5 and 10 and varying the number of QoS constraints between 2 and 5. Figure 5 (b) shows that the optimality of our algorithm is generally satisfactory (more than 90%) and it can reach 100%. Overall, both figures show that the optimality of our algorithm varies between 90% and 100% independently from the number of services and the number of QoS constraints, except for low service populations.

Comparing the optimality of QASSA to the optimality obtained in [16], for the same configuration (as both works use different number of activities and services) both works produce roughly the same optimality.

### 4.1.1 IMPACT OF THE AGGREGATION APPROACH

We propose to evaluate QASSA with respect to various QoS aggregation approaches, notably worst-case, best-case, and mean-value approach. We set the number of QoS constraints to 5, we vary the number of activities in the composition between 10 and 50, and we vary the number of services per activity between 50 and 200. Figure 6 depicts the execution time of QASSA associated with the worst-case, mean-value, and best-case aggregation approaches, respectively. We can clearly notice that aggregation approaches have no noteworthy effect on the execution time of QASSA, as they require nearly the same computational effort (they perform similar aggregation operations).

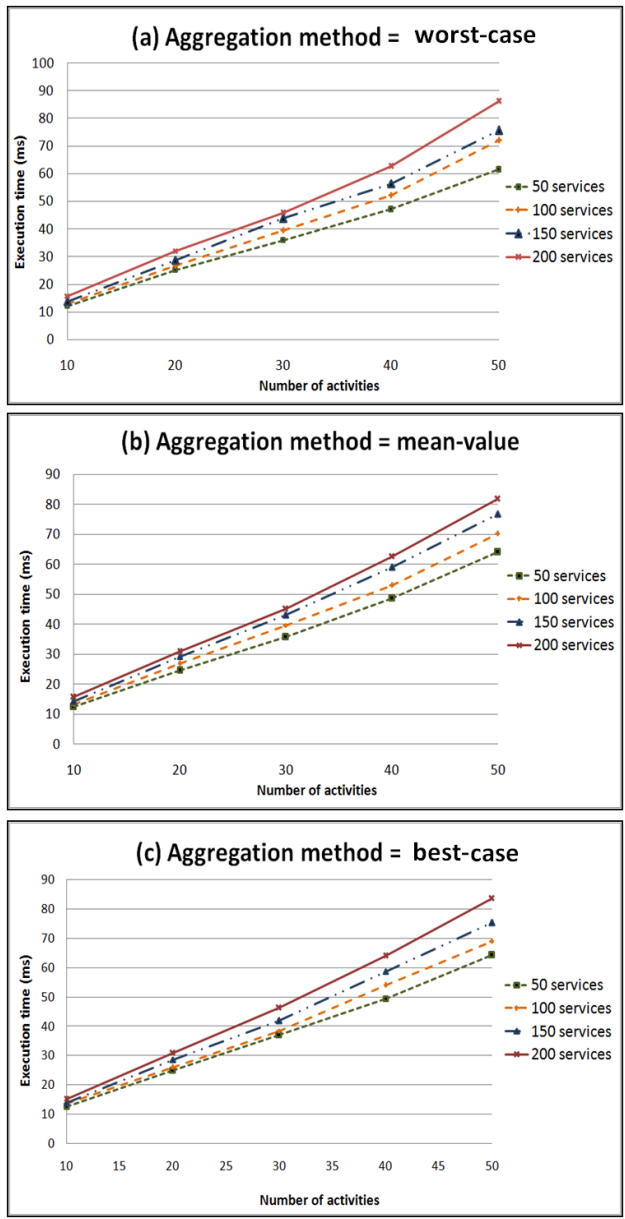

*Figure 6. Execution time wrt to the (a) worst-case, (b) best-case and (c) mean-value aggregation methods*

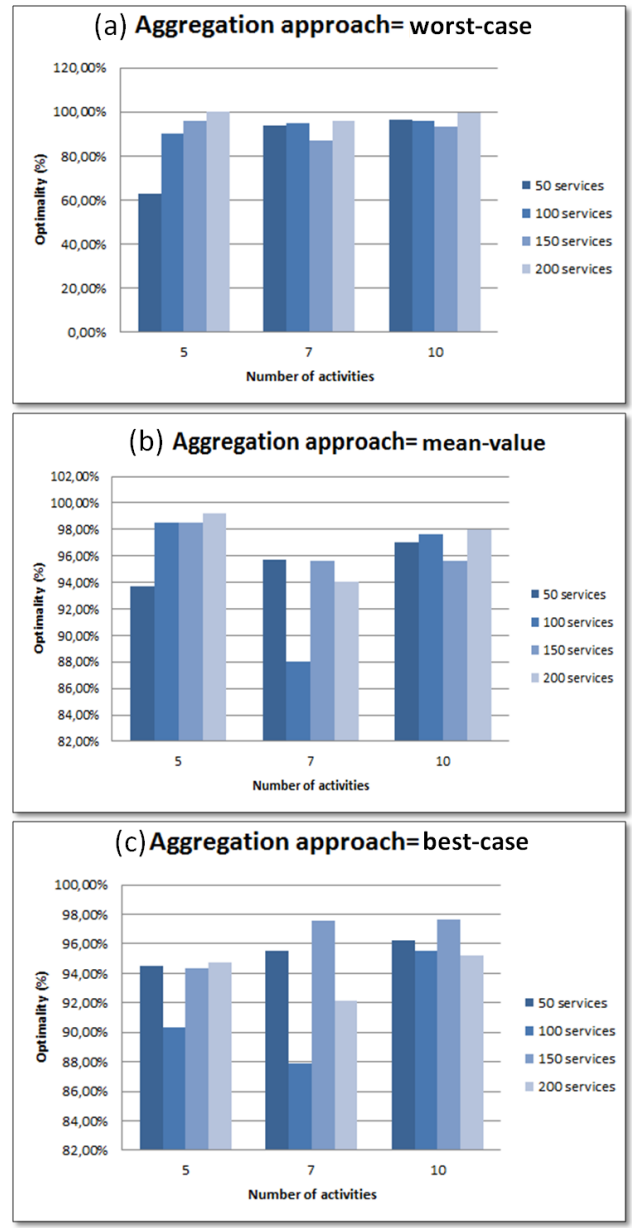

*Figure 7. Optimality of the algorithm wrt to the (a) worst case, (b) best-case, and (c) mean-value aggregation methods*

Concerning the optimality of QASSA, we measure it while decreasing the number of activities (we vary them between 5 and 10) in order to reduce the size of the problem, hence obtaining the optimal QoS promptly. Figure 7 depicts the optimality of QASSA associated with the (a) worst-case, (b) mean-value, and (c) best-case aggregation approaches. The figure shows that the optimality of QASSA slightly decreases from (a) to (b) and from (b) to (c). The best optimality is associated with the worst-case aggregation approach (it reaches 100%); for the mean-value aggregation approach the optimality does not exceed 99%; whereas for the best-case aggregation approach it is limited to 97%. This can be explained by the fact that when QASSA is more stringent and considers the worst QoS values, it discards service compositions with lower QoS and keeps only those closer to optimal.

### 4.1.2 THE IMPACT OF QOS REQUIREMENTS

The degree to which users are demanding (i.e., how strict are their QoS requirements) obviously impacts the number of service compositions able to meet these requirements. Accordingly, we propose to evaluate QASSA with respect to various values of the global QoS requirements imposed on the user task. In practise, determining such requirements is not trivial and requires real-world scenarios (i.e., real requirements), as well as it depends on the user profile (e.g., whether users are demanding or not). Existing QoS-aware service selection algorithms do not give a systematic method for setting meaningful global QoS requirements. To cope with this issue, we opt for a statistical approach that allows for determining global QoS requirements based on QoS values qi of service candidates. Specifically, we set the user global QoS requirement $u_i$ (associated with the QoS property pi) to two values:

$$u_i = \begin{cases} Agg(m) \\ Agg(m+\sigma) \end{cases} \quad (3)$$

where *Agg* is a function aggregating QoS values of services with respect to the structure of the composition, m and $\sigma$ are respectively the mean value and standard deviation of QoS values $q_i$ of candidate services associated with each activity. As the QWS dataset deals with a large number of services, the central limit theorem [22] states that the underlying QoS values follow the normal distribution law. Thus, setting the values m and $\sigma$ as local constraints allows for discarding respectively, 50% and 84,1% of service candidates associated with each activity in the user task.

Figure 8 depicts the execution time of QASSA associated with the global QoS requirements set to *m* and $m + \sigma$, respectively. In these figures, we notice that the execution time is roughly the same for both values of the global QoS requirements, which can be explained by the fact that these requirements are considered only during the global selection phase. That is, the local selection phase produces the same result (for a given set of services) independently from the global QoS requirements, and since the local selection is highly selective (i.e., it produces few services), the execution time of the global selection would be nearly the same for different values of the global QoS requirements.

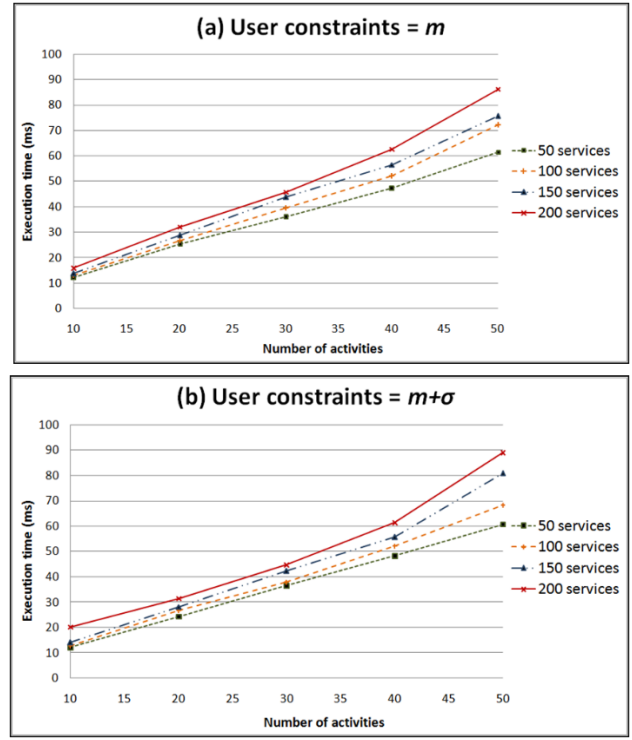

*Figure 8. Execution time while fixing global QoS requirements to (a) m, and (b) m+ $\sigma$*

Concerning the optimality of QASSA, Figure 9 depicts the obtained optimality results when the global QoS requirements are respectively set to *Agg(m)* and *Agg(m+ $\sigma$ )*. This figure shows that, the optimality produced by QASSA considerably decreases when QoS requirements are set m + $\sigma$, notably for a low number of activities (5 and 7 activities) and a low number of services per activity (less than 150 services). This can be explained by the fact that QASSA discards more service compositions when the values of global QoS requirements increase, hence the probability to find a service composition with high optimality decreases, particularly for a reduced number of activities and services per activity.

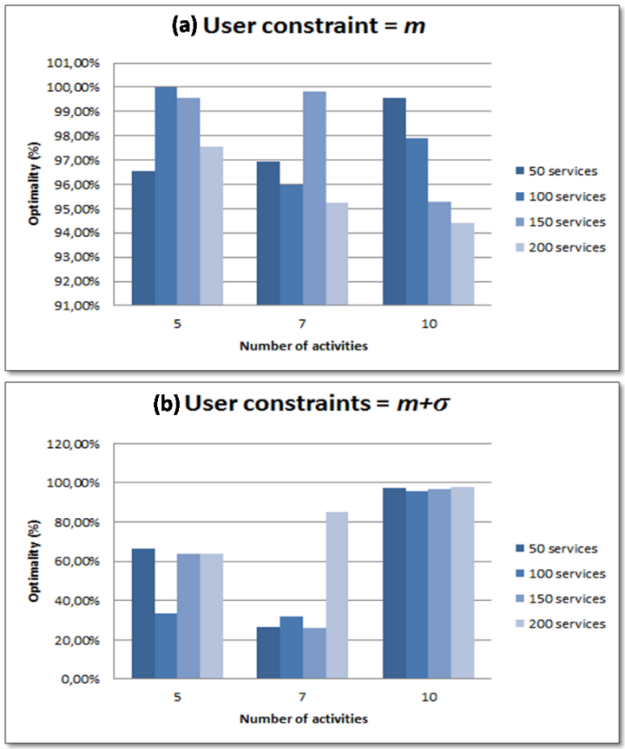

*Figure 9. Optimality of the algorithm while fixing global QoS requirements to (a) $m$, (b) $m + \sigma$*

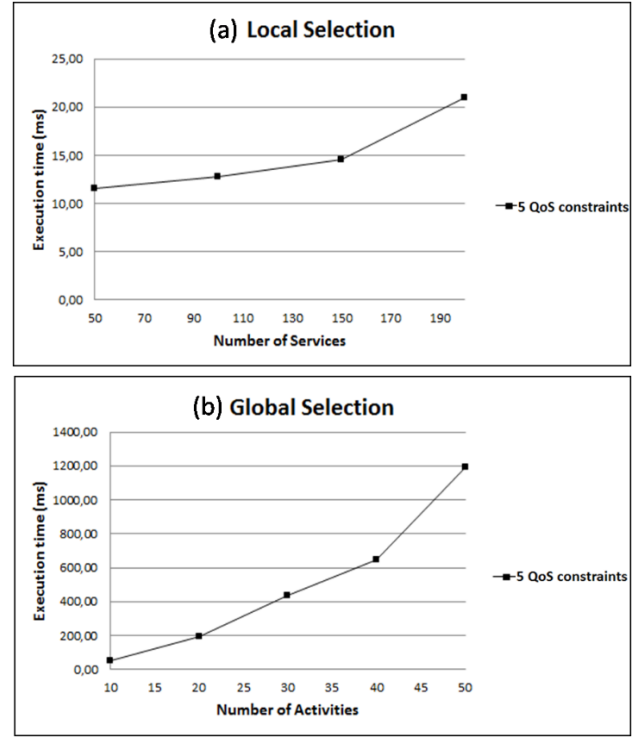

*Figure 10. Execution time of the (a) local selection and (b) global selection of distributed QASSA*

## 4.2 Performance of QASSA (the distributed version)

We evaluate the distributed version of QASSA using the experimental setup detailed in the right column of Table 1. The distributed design of QASSA changes two main factors compared with the centralised version, notably: (i) the communication cost between the devices participating in fulfilling the user task, and (ii) the hardware setup underpinning the execution of the algorithm. Both features do not impact the optimality of QASSA, thus the following experiments focus only on the execution time metric. Additionally, we assume that the communication cost is negligible compared to the overall execution time of the algorithm (further details about the network delay can be found in, e.g., [23]). Thus, the execution time presented in these experiments concerns only the local and global selection phases of the distributed version of QASSA (see Figure 10).

For the local selection, the execution time is measured for only one activity (indeed, each helper device processes in parallel local selection for a single activity in the user task). We fix the number of QoS constraints to 5 and vary the number of services between 50 and 200. Whereas, for the global selection, we fix the number of QoS constraints to 5, the number of services to 200, and we vary the number of activities in the user task between 10 and 50. Despite the relatively limited hardware resources used in these experiments, QASSA shows satisfactory timeliness with respect to on-the-fly service composition in ubiquitous environments. Indeed, the local selection is executed in at most 25 ms, whereas the global selection is executed in at most 1.2 s, thus the overall algorithm can be accomplished in less than 1.5s, depending on the size of the user task and the number of services per activity [21].

## 5 Related Work

Surveying QoS-aware service selection algorithms represents a broad topic. In this section, we focus on algorithms in line with the trend of combining local and global selection techniques, as it represents a general and powerful technique to extract optimal compositions in diverse scenarios [3]. In this context, Alrifai et al. [24] present a selection algorithm that starts from the global level and resolves the selection problem at the local level. The authors proceed by decomposing global QoS constraints (i.e., imposed by the user on the whole composition) into a set of local constraints (i.e., for individual sub-tasks, parts of the whole composition). To do so, they use MILP techniques to find the best decomposition of global QoS constraints. The main drawback of this approach is that it relies on a greedy method for the decomposition of QoS constraints, which produces strict local QoS constraints that may discard a lot of service candidates. In a more recent work [16], the authors attempt to enhance their approach by relaxing the local QoS constraints as much as possible while not violating the global constraints. While their recent approach may improve the obtained results, conceptually it does not resolve the problem of discriminating potential good service candidates.

The same authors present another approach [25] combining local and global selection techniques, but in another way. The authors start by the local selection phase. They use two techniques to reduce the number of services investigated for each activity in the user task. First, they use the *skyline* concept [26] as a technique to determine the most interesting services in terms of QoS. Once skyline services are determined, the authors cluster them into several clusters using K-means, and then they select a representative service for each cluster. At the global level, the authors compose the representative services selected at the local level, and check whether the composition meets global QoS requirements using MILP. This approach also presents several drawbacks. Concerning the algorithm itself, the authors claim finding the optimal service composition, because they assume that skyline services are the best services in terms of QoS, which is not true. Indeed, a skyline service is a service that has the highest (i.e., the best) value for one or more QoS properties, whereas for the remaining QoS properties it may have very low values. Regarding this definition, it is possible that a non-skyline service with high values (and not the highest) for all QoS properties yields a higher overall QoS than a skyline service. Concerning the performance of the algorithm, during the local selection phase the authors execute K-means $Z.(T/2)$ times (where Z is the number of activities in the user task, and T is the number of service candidates investigated for a given activity), which represents a high number of iterations, especially when it deals with a large number of service candidates. In our approach, we execute K-means++ (which already outperforms K-means) $Z.N$ times where N is the number of QoS properties. The complexity of our local selection phase is then reduced compared to [25], since the number of QoS properties is always limited compared to the number of service candidates. Additionally, at the global selection phase, the authors execute MILP iteratively until a near-optimal composition is found. In each iteration, the set of representative services with the highest QoS utilities is investigated. This approach may end up executing MILP $α$ times, where $α$ is the number of representative services, which means also a high number of iterations when it deals with a large number of representative services.

Another approach combining local and global selection techniques is presented by [19]. Similar to [24], the authors decompose global QoS constraints into local constraints using MILP. Based on the local QoS constraints, they select services for each activity in the user task. Then, they compose the locally selected services and check whether the composition meets global QoS constraints, using MILP again. The main advantage of this approach is that it executes local selection in a distributed way similarly to our approach. However, they decompose the global QoS constraint imposed on a given QoS property into the average values of that property associated with the services of each activity, which is not accurate and may discriminate a number of service candidates. A similar approach is presented by Jin et al. [27]. The authors decompose global

QoS constraints into local constraints using MILP, then they perform local selection. The main shortcoming of this approach is that it does not guarantee meeting global QoS requirements.

An interesting approach is presented by Liu et al. [28]. The authors propose a QoS-aware service selection algorithm which also combines local and global selection techniques. They use the *convex hull* concept [29] as a local selection technique. At the global level, the authors randomly establish an initial composition, and they try to enhance it using services selected by the convex hull. The main drawback of this approach is that it closely depends on the initial composition. Recently, Rodriguez-Mier et al. [3] introduce a QoS-aware service composition algorithm that combines local and global selection techniques, while considering service dependencies (input/output and precondition/effect dependencies). The introduced algorithm allows finding the optimal service composition. However, it considers only a single QoS objective.

Parejo et al. [30] introduce a hybrid QoS-aware service composition solution that combines two metaheuristics, viz., Greedy Randomised Adaptive Search Procedure (GRASP) and Path Relinking (PR). In this approach, GRASP is used for initialising the set of services to be investigated by the PR algorithm, which selects near-optimal compositions. This approach evaluates service compositions using a global utility function, which reduces multi-objective QoS-aware service composition to a single objective optimisation, thus leading to the issue of balancing low values of one or more QoS properties by good values of other properties.

To enable fine-grained management of QoS trade-offs, Chen et al. [7] introduce a Pareto set model for QoS-aware service composition. The introduced model defines multi-objective QoS dominance relationships between service candidates, as well as between service compositions. Based on this model, the authors proceed, at first, through a local selection phase that prunes service candidates by dominance relationships and the validation of user QoS constraints at the local level. After that, a global selection phase is performed based on dominance relationships between feasible service compositions. By considering only Pareto-optimal services, the proposed approach may discard prominent solutions. Indeed, the set of feasible Pareto-optimal service compositions may not be composed of the

Pareto-optimal services [8]. That is why in QASSA, we do not discard services with high QoS (even if they are Pareto-dominated by other services), we rather consider clusters of services having roughly the same high QoS, and we provide a flexible mechanism to decide about the trade-offs between QoS objectives.

## 6 CONCLUSIONS

This paper introduces QASSA, a QoS-aware service selection algorithm for ubiquitous computing environments.

QASSA defines service selection under global QoS requirements as a set-based optimisation problem, and solves this problem by combining local and global selection techniques. It introduces a novel method that uses clustering techniques to enable fine-grained management of trade-offs between QoS objectives. Moreover, QASSA considers jointly: (i) service dependencies, (ii) adaptation at run-time, and (iii) both centralised and distributed design fashions.

In practise, QASSA shows satisfactory timeliness and optimality, hence representing an efficient mean to achieve complex user tasks in ubiquitous environments.